\documentclass[letterpaper, 10pt, conference]{ieeeconf}                                                              
\IEEEoverridecommandlockouts                           
\usepackage{microtype}
\usepackage{verbatim}
\usepackage[singlespacing]{setspace}
\usepackage{hyperref}
\usepackage{balance}
\usepackage{multirow}
\usepackage{graphics}
\usepackage{trimclip} 
\usepackage{algorithm}
\usepackage{algorithmicx,algpseudocode}

\newtheorem{definition}{Definition}
\newtheorem{remark}{Remark}

\usepackage{xr}
\usepackage{hyperref}
\usepackage{url}
\usepackage{amsmath}
\usepackage{amssymb}
\usepackage{dsfont}
\usepackage{upgreek}
\usepackage{bbm}			
\usepackage{graphicx}
\usepackage{epstopdf}
\usepackage{tikz}
\usepackage{float}
\usepackage{booktabs} 
\usepackage{subfigure}
\usepackage{xspace}
\usepackage[scaled=.7]{beramono}

\newcommand{\DM}{administrator\xspace}

\newcommand{\SPfull}{Secretary Problem\xspace}

\newcommand{\SSPfull}{Sequential Selection Problem\xspace}
\newcommand{\SSP}{SSP\xspace}

\newcommand{\WSSP}{WSSP\xspace}
\newcommand{\WSSPfull}{Warm-starting Problem\xspace}
\newcommand{\DRA}   {DRA\xspace}
\newcommand{\DRAfull}{Dynamic Resource Allocation\xspace}
\newcommand{\RDRAfull}{Restricted Dynamic Resource Allocation\xspace}
\newcommand{\RDRAshort}{Restricted DRA\xspace}
\newcommand{\RDRA}{RDRA\xspace}
\newcommand{\MSSPfull}{Multi-round Sequential Selection Process\xspace}
\newcommand{\MSSP}{MSSP\xspace}
\newcommand{\DP}{DP\xspace}
\newcommand{\DPs}{DPs\xspace}

\newcommand{\SDRAfull} {Sequential Dynamic Resource Allocation\xspace}
\newcommand{\SDRAshort} {Sequential DRA\xspace}
\newcommand{\SDRA}   {SDRA\xspace}

\newcommand{\CCM}   {CCM\xspace}
\newcommand{\CCMstar}   {$\text{CCM}^*$\xspace}
\newcommand{\CCMfull}   {Cutoff-based Cost Minimization\xspace}
\newcommand{\MEAN}   {MEAN\xspace}
\newcommand{\MEDIAN}   {MEDIAN\xspace}

\renewcommand{\t}		{k}
\newcommand{\ninf}[1]	{\text{sum}(\Xbold{})}
\newcommand{\spaceInfection} 	{\{0,1\}}	
\newcommand{\cost}    		 {\phi}

\newcommand{\policybold}   		 {\mathbf{\Pi}}
\newcommand{\info}    		 {\mathcal{I}}
\newcommand{\nodes}    		 {\mathcal{V}}

\newcommand{\cand}				 {}

\newcommand{\Xbold}[1]	               {\mathbf{X}_{#1}} 
\newcommand{\val}			      {S}
\newcommand{\Sbold}	               {\mathbf{\val}_{\cand}}

\newcommand{\symb}[1]		   {#1^R}

\newcommand{\Cpres}	       		{\symb{C}}

\newcommand{\Abold}		        { \mathbf{R}_{\cand} }
\newcommand{\A}[1]				{R_{#1}}

\newcommand{\Rbold}[1]    {\mathbf{R}_{#1}}

\newcommand{\C}[2]    {C_{#1,#2}}
\newcommand{\Co}[1]    {\Cpres_{#1}}
\newcommand{\Cbold}[1]    {\mathbf{C}_{#1}}
\newcommand{\Cobold}[1]    {\symb{\mathbf{C}}_{#1}}
\newcommand{\Rankset}[2] 			{\mathcal{P}_{#1}(#2)}
\newcommand{\Graph}    {\mathcal{G}}

\makeatletter
\g@addto@macro\bfseries{\boldmath}
\makeatother
\newcounter{phase}[algorithm]
\newlength{\phaserulewidth}
\newcommand{\setphaserulewidth}{\setlength{\phaserulewidth}}

\makeatother
\setphaserulewidth{.3pt}

\newcommand{\Sec}[1]		{Sec.\,\ref{#1}}
\newcommand{\Fig}[1]		{Fig.\,\ref{#1}}

\newcommand{\Alg}[1]		{Alg.\,\ref{#1}}

\newcommand{\Definition}[1]{Definition~\ref{#1}}

\newcommand{\vs}   			{vs.\@\xspace}
\newcommand{\ie}   			{i.e.\@\xspace}
\newcommand{\eg}   			{e.g.\@\xspace}
\newcommand{\etc}   		{etc.\xspace}
\newcommand{\wrt}   		{w.r.t.\@\xspace}

\newcommand{\st}   			{\mbox{s.t.}\xspace}

\newcommand{\Exp}[1]    {\mathbb{E}[#1]}
\newcommand{\Prob}      {\mathbb{P}}

\newcommand{\real}     {\mathbb{R}}

\newcommand{\natnostar} {\mathbb{N}}
\newcommand{\nat}     {\mathbb{N}^*}
\newcommand{\pdf}				{p.d.f.\xspace}

\newcommand{\mydots} 	{...}
\newcommand{\algComment}[1] 	{\hfill/\!/\,{#1}}

\newcommand{\inlinetitle}[2]  {\vspace{4pt}\noindent\textbf{\emph{#1}{#2}}}

\renewcommand*{\top}{{\mkern-1.5mu\mathsf{T}}}

\newcommand{\obullet}{\square}

\newcounter{marginNoteCounter}

\begin{document}
\title{\LARGE \bf \SDRAfull for Epidemic Control}
\date{}
\author{Mathilde Fekom \quad Nicolas Vayatis \quad  Argyris Kalogeratos
\thanks{\!\!\!\!\!\!$\obullet$~The authors are with CMLA -- ENS Paris-Saclay, 94230 Cachan, France. Emails:~
{\tt\footnotesize  \{fekom,\,vayatis,\,kalogeratos\}@cmla.ens-cachan.fr}.}
\thanks{\!\!\!\!\!\!$\obullet$~Part of this work was funded by the French Railway Company, SNCF, and the IdAML Chair hosted at ENS Paris-Saclay.
}
}

\maketitle
\begin{abstract} Under the \emph{Dynamic Resource Allocation} (DRA) model, an \DM has the mission to allocate dynamically a limited budget of resources to the nodes of a network in order to reduce a diffusion process (\DP) (\eg an epidemic). The standard DRA assumes that the \DM has constantly \emph{full information} and \emph{instantaneous access} to the entire network. Towards bringing such strategies closer to real-life constraints, we first present the \emph{\RDRAshort} model extension where, at each intervention round, the access is restricted to only a fraction of the network nodes, called \emph{sample}. 
Then, inspired by sequential selection problems such as the well-known Secretary Problem, we propose the \emph{\SDRAshort} (\SDRA) model. Our model introduces a sequential aspect in the decision process over the sample of each round, 
offering a completely new perspective to the dynamic \DP control. Finally, we incorporate several sequential selection algorithms to \SDRA control strategies and compare their performance in SIS epidemic simulations. 
\end{abstract}

\section{Introduction}\label{sec:intro}
Compartmental models have been extensively studied in epidemiology since early last century. In recent years, they have gained much wider attention due to their simple analytic formulations that can model modern problems related to information diffusion and social epidemics, \eg rumor spreading \cite{rumorSpreading2018} and other social contagions \cite{SISa_obesity}. Being able to control efficiently undesired diffusion processes (\DPs) is very crucial for public health and security. Yet, it is a difficult problem that in fact gets instantly much more complicated the moment one starts including more realistic constraints or objectives. This explains why most studies of the literature, despite providing high-level insights about the phenomena, remain rather far from being applicable in practice.

A source of limitations is the theoretical \emph{interaction model} one considers, along with its network-wise abstraction level (\eg macro- vs microscopic modeling), which may be over-simplistic for the analyzed phenomenon. 
Another source of shortcomings is the requirement for having information regarding the \emph{system state}, such as the infection state of nodes or the network connectivity. 
Finally, limitations come from the way a control model assumes it can intervene to the \DP, in a static or dynamic fashion to the evolution to the process, and using certain kinds of resources or actions.

\emph{Dynamic Resource Allocation} (DRA) \cite{Scaman15, Scaman16} is a model for network control, originally developed for SIS-like processes \cite{Mieghem09} (the nodes are either infected or healthy without permanent immunity) that distributes a limited budget of available treatment resources on infected nodes in order to speed-up their recovery. The resources are non-cumulable at nodes (\ie each node gets at most one resource) and cannot be stored through time. The \emph{score-based} DRA formulation introduces an elegant way, through a simple score value, of assessing the criticality of each node individually for the containment of the DP. Then, the \DM only has to ensure that at each moment the resources will be spent on the infected nodes with the highest scores. Among the proposed options, a simple yet efficient local score is the \emph{Largest Reduction in Infectious Edges} (LRIE) \cite{Scaman15}, which depends on the infection state of the neighbors, hence it needs to be updated regularly during the process. 

The motivation of our work is to bring the score-based DRA modeling closer to reality. In the majority of real-life scenarios, authorities have access to limited information regarding the network state, and can reach a limited part of the population to apply control actions (\eg deliver treatments). Even more importantly, the decision making process is essentially a sequence of time-sensitive decisions over choices that appear and remain available to the \DM only for short time, also with little or no margin of revocation. 
An intuitive paradigm to consider is how a healthcare unit works: patients arrive one-by-one seeking for care, and online decisions try to assign the limited available resources (\eg medical experts, beds, treatments) to the most important medical cases \cite{Bekker17, Kabene06, Gnanlet09}.
By establishing a link between the \DRA problem and the sequential decision making literature, our work offers a completely new perceptive to dynamic \DP control. 
Among the existing \emph{\SSPfull}\emph{s} (\SSP) that have been widely studied, the most well-known is the \SPfull \cite{Ferguson89}. Our aim, however, is to propose a concordant match to the \DP control setting described above.

Concerning our technical contribution, we first present the \emph{\RDRAshort} (\RDRA) model, in which each time the \DM can decide the reallocation of the resources only among a random sample of currently reachable nodes.
On top of that, we next propose the special case of \emph{\SDRAshort} (\SDRA) where the latter sample of nodes is provided with a random \emph{arrival order}, forcing the \DM to decide for the resource reallocation sequentially according to the characteristics of the incoming nodes. 
We believe that the major achievement of our modeling is that it manages to create a new playground where \SSP algorithms can be incorporated to the \DP control, and this way makes control strategies more applicable in real conditions. The implementation of existing online algorithms such as the \emph{hiring-above-the-mean} \cite{Broder09} or even the more effective \emph{\CCMfull} \cite{Fekom19}, leads to \SDRA strategies that manage to reduce the \DP in a comparable fashion to the unrestricted \DRA.
\newpage

\section{The Sequential \DRA}\label{sec:setting}
\subsection{Setting and scoring function}
Consider the environment set by a fixed network represented by the graph $\Graph(\nodes,\mathcal{E})$ of $|\nodes| = N$ nodes and $|\mathcal{E}| = E$ edges. To simplify the presentation we directly suppose that the \DP that takes place is a continuous-time Markov process \cite{Mieghem09}, so that at each time instance $t\in \real_+$ there can be at most one event of node state change in the network. In particular, we consider an SIS-like epidemic, where nodes are either healthy (susceptible to infection), or infected. The infection state of the network is denoted by $\mathbf{X}_{t}=(X_{1,t},\mydots,X_{N,t})^\top \in \spaceInfection^N$ \st $X_{i,t}=1$ if node $i \in \nodes$ is infected and $0$ otherwise. The infection spreads from each infected node to its healthy neighbors. Nodes are equipped with self-recovery but they never achieve permanent immunity. 

An \emph{\DM} has the mission to reduce the DP by managing a fixed budget of $b \in \nat$, $b\ll N$, resources that help the receiving nodes leaning towards the healthy state. The resources are regarded as reusable treatments that cannot be stored through time and are non-cumulable at nodes (\ie at most one on each node). 

The \emph{\DRAfull} (DRA) \cite{Scaman15, Scaman16} dynamically determines the resource allocation vector $\Rbold{t}=(R_{1,t},\mydots,R_{N,t})^\top \in \{0,1\}^N$ where $R_{i,t} =1$ if a treatment is allocated to node $i$ at time $t$ and $0$ otherwise; subject to $\sum_i R_{i,t} = b,\, \  \forall t \in \real_+$. The strategy is dynamic and adapts to the infection state. The stochastic transition rate from state $x$ to state $y$ at time $t$ is given by $p_t(x, y)\in \real$ and depends on the allocation indicated by $\Rbold{t}$. Finally $\mathbf{p}_t(x) = (p_t(x, y_1),\mydots,p_t(x, y_{2^N}))^\top \in \real^{2^N}$ is the vector with the probabilities to go from the system state $x$ to every other possible state at time $t$.

The \emph{score-based DRA} assumes that there exists a scoring function $s: \nodes \rightarrow \real$ that computes a score $S_{i,t}$ for each node $i$ at time $t$ according to the mission. At any given moment, the nodes with the highest scores are those to receive the resources. This class of strategies depends on the efficiency and the size of the available budget of resources, and also on the efficiency of the scoring function in indicating the most critical nodes. 

\begin{figure}[t]
\centering
\includegraphics[width=1\linewidth, viewport=10 15 680 665,clip]{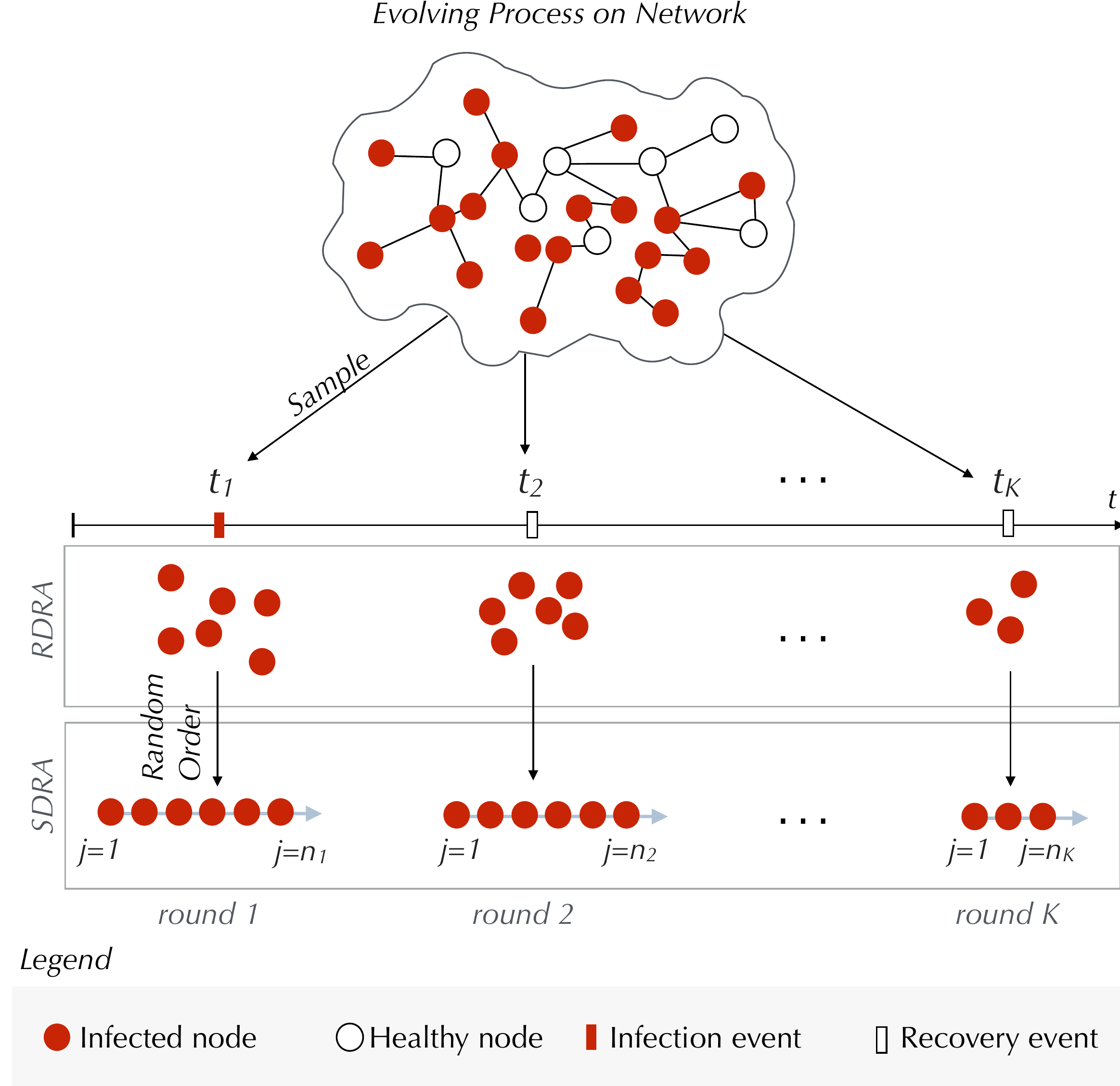} 
\vspace{-1.5em}
\caption{The sequential evaluation of candidates in the \SDRA model.}
\label{fig:time_scale}
\end{figure}

\subsection{\RDRAfull}
The standard \DRA strategies are build on the strong assumption whereby the \DM has always full information and access to the network, which is apparently infeasible in many practical cases. To relax this requirement we introduce the \emph{\RDRAshort} (\RDRA) model in which only a fraction of nodes are reachable at each moment. We work with two reasonable assumptions: 1)~the access to nodes and the information acquisition about them are regarded as inextricable, and 2)~the set of treated nodes, $\Cobold{t} = \{i \in \nodes : R_{i,t} = 1\} \subset \nodes, \, \ \forall t \in \real_+$, is always accessible.

\begin{definition}{\emph{\RDRAfull} (\RDRA) strategy $\policybold_{t}(\info)$} is a DRA strategy that includes the number of resources $b\in \nat$ and the scoring function $s:\nodes\rightarrow \real$. At any moment, the \DM has access to nodes in the set $\info \subset \nodes$, in addition to those currently treated, $\Cobold{t}\subset \nodes$. The strategy outputs a resource allocation vector, \ie $\policybold_{t} = \Rbold{t}, \, \forall t \in \real_+$.
\label{def:rdra}
\end{definition}

The default is the accessible set to be $\info = \Cbold{t}$, where $\Cbold{t}$ is called \emph{sample} and is defined below. Choosing to define $\info$ otherwise, results in special \RDRA cases.

\begin{definition}{\emph{Node sample} $\Cbold{t} $} is the set of accessible infected nodes at time $t$, $\Cbold{t} =(\C{1}{t},\mydots,\C{n_t}{t}) \subset \nodes$. Its size $n_t=f(\Xbold{t}) \in \nat$ is given by $f:\spaceInfection^N \rightarrow \nat$ that is a function of the infection state. 
The probability of observing a sample $c \subset \nodes$, given its size $n\in \nat$ and the network state $x\in \spaceInfection^N$, is $\Lambda_{t}(c;n,x)=\Prob(\Cbold{t}=c \mid |\Cbold{t}| = n, \Xbold{t} = x)$. In short we write $\Cbold{t} \sim \Lambda_{t}(n,\Xbold{t})$. 
\label{def:sample}
\end{definition}

\subsection{\SDRAfull}\label{sec:time_scale}
The \RDRA's assumption of having simultaneous access to all the nodes of a sample remains far from being realistic. Refining further the access constraints, we present the \emph{\SDRAshort} (\SDRA) model that is enriched with a phase of sequential processing of the sample. Same as in the standard DRA, the resource allocation is questioned whenever there is a change in the infection state of the network. This defines what we call a \emph{round} of allocation decisions.

\begin{definition}{\emph{Round} $k$} is an event of (re)-allocation of resources on the network. The series of rounds is defined by the sequence of time instances $(t_k)\in\real^K$ characterized by the recurrence: 
\begin{equation}
t_{k} = t_{k-1}+ \text{min}(\delta t \mid ||\Xbold{t+\delta t} - \Xbold{t} ||=1),\, \ \forall k \leq K,
\end{equation}
where $t_{0} = 0$, and $K$ is the total number of rounds. 
\label{def:round}
\end{definition}

Note that by definition a round, overall, acts at a much smaller time-scale compared to the \DP. Moreover, the round duration is further divided into $n_\t$ time intervals. The purpose of introducing the concept of round is to make the reallocation sequential at the time-scale of the round duration, that is to create an order in which nodes of the sample are evaluated for a treatment. This replaces the way of reassigning altogether the $b$ treatments, which is used by \DRA or \RDRA strategies. Hence, the discrete index $j \in \{1,\mydots,n_k\}$ characterizes the sequential arrival order of candidates, \eg $j=1$ and $j=n_k$ are respectively the first and last candidates of round $k$.

Since a round is a measure of time, each variable can be defined by its value within a round, for instance we write $\Xbold{k}$ for the infection state at round $k$, \ie at time instance $t_k$. Also, the \DM gains access sequentially on incoming candidates, therefore the variables might depend on the index $j \in \{1,\mydots,n_k\}$.

\begin{definition}{\emph{\SDRAshort} (\SDRA) $\policybold_{k}(\info_j)$} is the \RDRA strategy defined at time instances $(t_k)\in \real_+$, and where $\info_j=C_{j}, \, \forall j \leq n_k$, providing a uniformly random arrival order to the nodes of the sample.
\label{def:sdra}
\end{definition}

\begin{algorithm}[t]
\footnotesize
\caption{DP control with \RDRAshort}
{\bf Input:} $N$: population size; $b$: budget of resources; $\Xbold{0}$: initial infection state; $\mathbf{p} ({x})$: transition probability from state $x$ to every other state; $f$: function that gives the number of accessible nodes; $\Lambda $: \pdf of the sample; $\policybold $: \RDRAshort strategy; \emph{isSequential}: specifies if the strategy is \RDRA (false) or \SDRA (true). 

{\bf Output:} $\Xbold{}$: final network state, $\Rbold{}$: final allocation of the resources
\begin{algorithmic}[1]
\State{$\Xbold{} \leftarrow \Xbold{0}$} \algComment{initialize the infection state}
\State{$\Rbold{}(\text{randp}(b,N)) \leftarrow 1$} \algComment{initialize the resource allocation}
\While{$\ninf{} \neq 0$}
\State {$n \leftarrow f(\Xbold{})$ } \algComment{compute the number of accessible nodes}
\State {$\Cobold{} \leftarrow \text{find}(\Rbold{} = 1)$} \algComment{currently treated nodes} 
\State {$\Cbold{} \sim \Lambda(n,\Xbold{})$ } \algComment{generate the sample} 
\If {\text{\emph{isSequential}} = true}
\For {$j =1\mydots n$}\algComment{loop of a selection round}
\State{$\Rbold{} \leftarrow \policybold(\Cobold{}, C_{j})$} \algComment{seq. update resource allocation}
\EndFor
\Else
\State{$\Rbold{} \leftarrow \policybold(\Cobold{},\Cbold{})$} \algComment{update resource allocation altogether}
\EndIf
\State{$\Xbold{} \leftarrow \mathbf{p}(\Xbold{})$} \algComment{update the infection state}
\EndWhile
\State {\Return $\Xbold{}$, $\Rbold{}$}
\end{algorithmic}
\label{alg:sdra}
\end{algorithm}

The way the \RDRA and \SDRA models operate is described in \Alg{alg:sdra}, an example is also depicted in \Fig{fig:time_scale}.

\section{From \DP control to a \MSSPfull}\label{sec:main}
\subsection{Link with the \SSPfull (\SSP)}
To the best of our knowledge, our work is the first to cast the dynamic \DP control as a problem where decisions are seen as in a \SSPfull (\SSP). 
Our purpose, though, is not to develop here a new \SSP but rather to resort to existing results of the field, and hence to connect the \SDRA problem described in \Sec{sec:setting} to a suitable \SSP framework. Features to consider are, for instance, to have single (or multiple) resource(s), finite (or infinite) horizon, \emph{score-based} (or \emph{rank-based}) objective function, \etc

It turns out that the biggest difficulty to find a match to existing methods is that our setting has a set of currently treated nodes when the selection round starts, as the $b$ resources should be constantly in effect in the network. Indeed, most SSPs consider a \emph{cold-starting} selection, \ie the \DM begins with an empty selection set. Here, each round gets \emph{initialized} since the $b$ currently treated nodes are already `selected'. Nonetheless, a \emph{warm-starting} variant has been investigated in \cite{Fekom19}. 

\begin{remark} Since the reallocation of resources takes place upon every change of the infection state of the network, at most one node can recover between two subsequent rounds.
\end{remark}

\begin{figure*}[t]
\centering
\clipbox{0pt 70pt 0pt 0pt}
{\includegraphics[width=0.55\linewidth, viewport=0 0 880 723,clip]{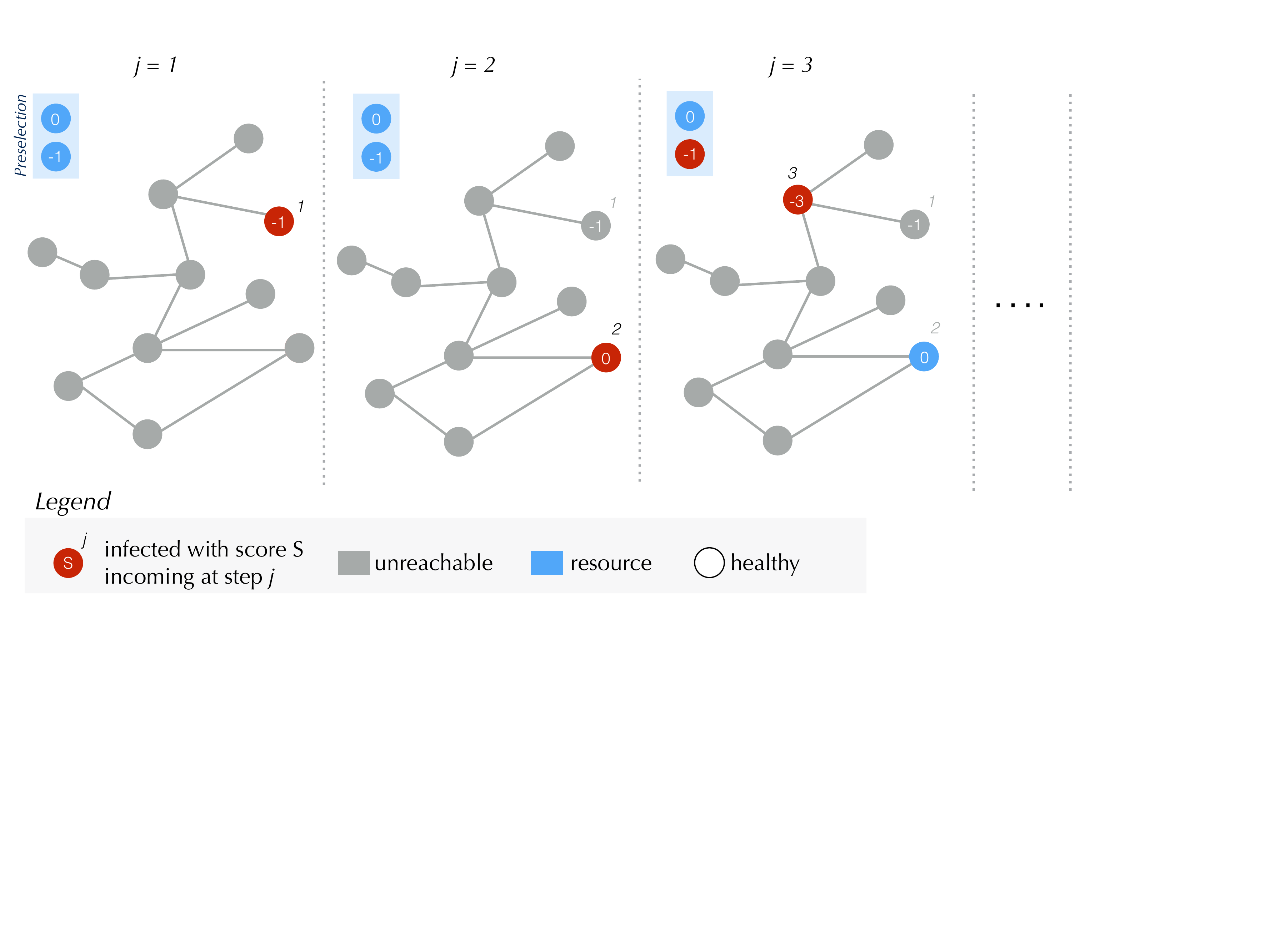}}
\clipbox{20pt 70pt 60pt 0pt}
{\includegraphics[width=0.55\linewidth, viewport=0 0 880 723,clip]{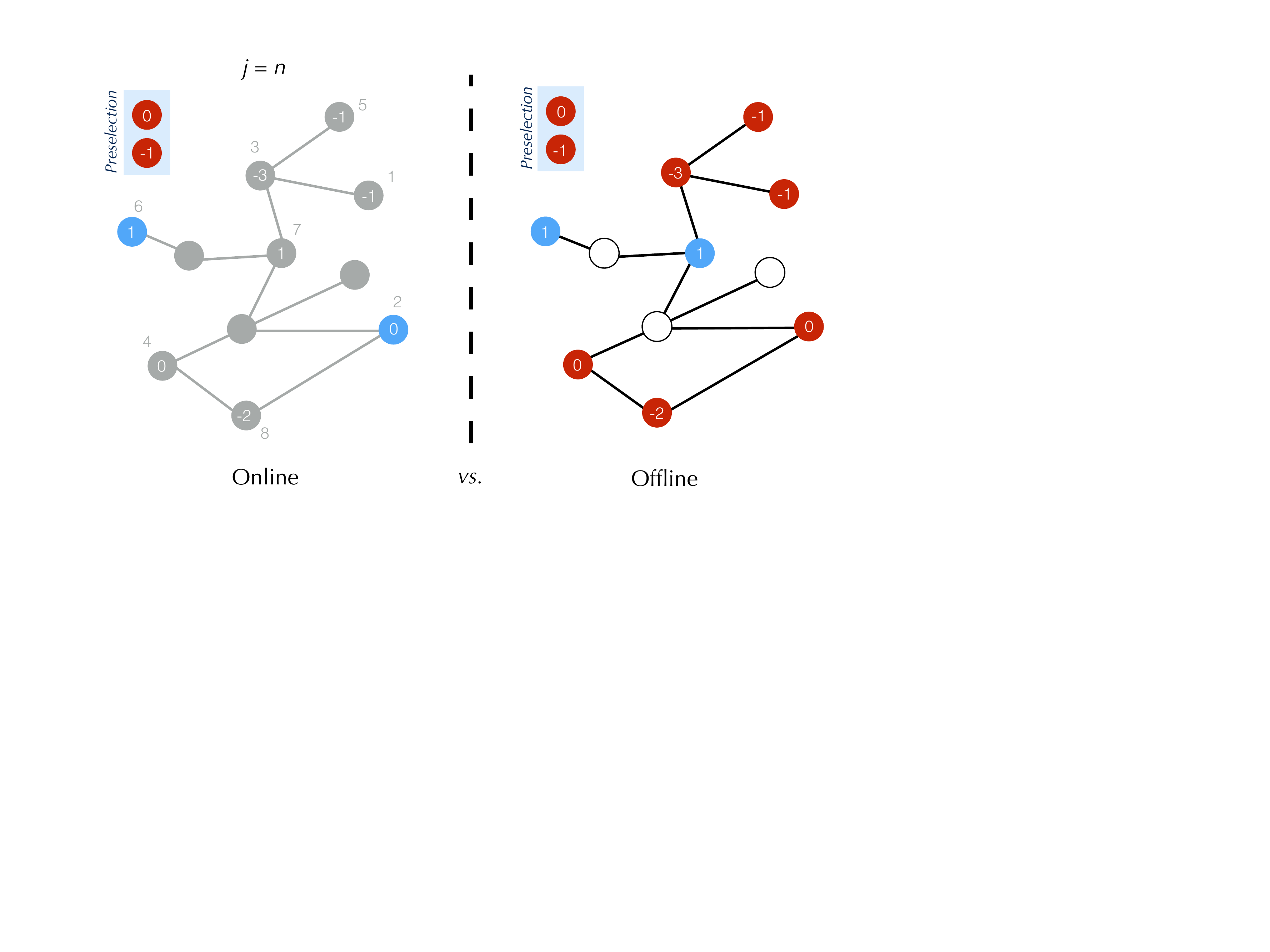}}
\vspace{-2.5em}
\caption{Example of the 3 first steps of a \SDRA round. Candidate nodes are sequentially incoming \wrt $j$ and possible reallocations are decided immediately; \eg at step $j=3$ a candidate is given a resource, that is withdrawn from a preselected. The result is compared to that of the RDRA (offline).}
\label{fig:online}
\end{figure*}

\subsection{\WSSPfull}\label{sec:mapping}

Inspired by \cite{Fekom19}, we map the problem of \DP control with \SDRA to a succession of separate \emph{\WSSPfull\!\!s} (\WSSP\!\!s), see \Definition{def:ssp}. Specifically, one round of the former corresponds to one instance of the latter. For convenience, within each \WSSP, the round subscript $k$ is dropped in our notations, \eg $\Cbold{k}$ becomes $\Cbold{}$. 

\begin{definition}{\emph{\WSSPfull} (\WSSP)} is an \SSP variant described by elements of different categories: 

\noindent1)~Background (included in $\mathcal{B}$) 

a)~Information
\begin{itemize}
\item $b \in \nat$: fixed budget of resources,
\item $s:\nodes \rightarrow \real$: scoring function \st $\Sbold = s(\nodes) \in \real^N$ is the score vector of the entire population,
\item $n\in \nat$: number of candidates to come.
\end{itemize}

b)~Initialization
\begin{itemize}
\item  $\Cobold{} = (\Co{1}{},\mydots,\Co{b}{}) \subset \nodes$: the subset of the population, called \emph{preselection}, to which resources are initially allocated when a round begins, \ie $\A{\Co{i}{}}{} = 1, \, \forall i \leq b$.
\end{itemize}

\noindent2)~Process \& Decisions
\begin{itemize}
\item $(C_{1},\mydots,C_{n}) \in \Rankset{n}{\nodes \backslash \Cobold{}}$: sequence of randomly incoming candidates for receiving a resource, where $\Rankset{l}{E}$ denotes the set of $l$-combinations of some finite set $E$,
\item $(\A{C_1}{}, \mydots, \A{C_n}{})\in \{0,1\}^n$: sequence of resource allocation decisions taken; giving a resource to a candidate immediately withdraws it from a preselected individual (recovered or not), \ie $\A{C_j}{}=1 \Rightarrow \exists\, i\leq b \  \st \ \A{\Co{i}{}}{} = 0$.
\end{itemize}
3)~Evaluation
\begin{itemize} 
\item The cost function is defined as:
\begin{equation} 
\cost_\mathcal{B} = \underset{R_i, i\in \mathcal{C} }{\max}(\Sbold \cdot \mathbf{R} )-(\Sbold \cdot \Abold) \, \in \real_+,
\label{eq:cost}
\end{equation}
where $\mathcal{C} = (\Cobold{}, C_{1},\mydots,C_n) \subset \nodes$.
\end{itemize}
\label{def:ssp}
\end{definition} 
The first term of the cost function defines the highest achievable score, while the second gives the score obtained from the sequential decisions. The sequence of incoming candidates being a random variable,
$\Exp{\cost_\mathcal{B}/b }$ is the objective function to maximize.

Two observations have to be made concerning the aforementioned mapping: 1)~it translates the objective of the DP control, \ie to minimize the percentage of infected nodes through time, into a \SSP objective: to minimize the expected cost function of the selected items, hence, $\eta_t = \Exp{\frac{1}{N}\sum_i X_{i,t}}$ is closely related to $\Exp{\cost_\mathcal{B}/b}$; 2)~it is done so that during each \WSSP instance, the \DM knows nothing about the infection state of the network, and simply selects online. Note that, when still having available resources (\eg allocated to preselected individuals that just recovered) while reaching the end of the sequence, then those resources are by default given to the last infected candidates to appear. 

\subsection{Offline \vs Online}

In our \DP context, a strategy is called \emph{offline} when it systematically selects the $b$-best reachable nodes and immediately assigns resources to them. As explained in the first section, the notion of `best' is given by an \emph{expert} through the scoring function $s:\nodes\rightarrow \real$, which prioritizes nodes according to their criticality for the spread of the \DP. The way this is achieved is not of this paper's concern, it is regarded as a `\emph{black box}'.
An \emph{online} strategy, however, can only examine candidate nodes one-by-one, see \Definition{def:sdra}. 

In \Fig{fig:online}, an example is displayed of the final resource allocation using an online and an offline strategy. Two resources are initially active, \ie $b=2$, represented by the blue nodes in the preselection. Scores are shown inside each infected node (the higher, the more critical). 
Consider for instance, an online strategy that gives a resource to each incoming node with a score higher than the average score of the preselection, here with scores $\{0,-1\}$. When the first candidate arrives ($j=1$) with a score of $-1$, it is not selected since it does not beat $-0.5$; however the second candidate ($j=2$) has a score of $0$, and a resource is reallocated to it from the worse preselected node. The score threshold to beat becomes $0$. The process continues, up to the last candidate ($j=n$). Here, the cost function is $\frac{1}{b}\cost_\mathcal{B} = (1+1)-(1+0)=1$, where the first term is the highest achievable average score (\ie the offline score), and the second term is the allocation resulting from the online strategy.
 
Ideally, an efficient Sequential \DRA strategy (online) should be as close as possible to the associated \RDRAshort strategy (offline), regardless the scoring function; in other words, having $\cost_\mathcal{B}$ as small as possible.

\subsection{Algorithms for \DP control}
The mapping we introduced in \Sec{sec:mapping} allows and suggests the implementation of online algorithms of the \SSP literature to sequentially control \DPs. In particular, the focus is put on two categories of online strategies: 
\begin{itemize}
\item \emph{cutoff-based}: it takes as input a given \emph{cutoff value} $c\in \natnostar$; it first rejects by default the first $c$ incoming candidates, called the \emph{learning phase} and then selects a candidate according to information gathered during the latter phase.
\item \emph{threshold-based}: a particular case of \emph{cutoff-based} strategies with $c=0$. A candidate is accepted if his score beats a specified \emph{acceptance threshold}.
\end{itemize}
We chose one indicative algorithm from each of the above classes, the \emph{Cutoff-based Cost Minimization} (\CCM) and the \emph{hiring-above-the-mean} (\MEAN); whose objectives are to minimize the expected sum of the ranks, or respectively sum of scores scores, of the treated nodes at the end of a round.

In \CCM, a notion of \emph{quality} $q \in ]0,1[$ of the preselected \wrt the sample has to be given as input. In this paper, we set $q_{k}=\phi_{\mathcal{B}, k-1}/b, \, \forall k\leq K$, where $\phi_{\mathcal{B},0}=0.5$. Then, a table $c^*(b,n,q) \in \natnostar^{b\times n}$ is computed by tracking the lowest point of the expected rank-based cost provided in \cite{Fekom19}. Finally, the acceptance threshold is essentially the $b$-th best score seen during the learning phase. For simplicity, we denote by \CCMstar the \CCM strategy with $c=c^*(b,n,q)$.

In \MEAN \cite{Broder09}, the acceptance threshold used is the average score of the preselection, thus it evolves with each selection of candidate. This strategy, although intuitive and easier to implement than \CCM, reaches its limit when the preselection is of poor quality with respect to the sample (and probably also with all the population of care-seekers). We also consider the strategy where the acceptance threshold used is their median score, \MEDIAN.

In the next section, both these types of algorithms are compared in various simulations. 

\section{Simulations}
\subsection{Experimental setup}
\inlinetitle{Network}{.}
The interactions among a population of $N$ individuals are modeled by a fixed, symmetric (undirected), and unweighted network with adjacency matrix $\mathbf{M} \in \{0,1\}^{N\times N}$. For each entry it holds $M_{ij}=M_{ji}=1$ if nodes $i$ and $j$ are linked with an edge, or $0$ otherwise. The connectivity structure is generated according to either a \emph{scale-free} (SF) or a \emph{small-world} (SW) network model.  The characteristic of the SF type is that its node degree distribution follows a power law, hence few nodes are \emph{hubs} and have much more edges than the rest. We use the Barab\'asi-Albert \emph{preferential attachment} model \cite{Barabasi99} that  starts with two connected nodes and, thereafter, connects each new node to $m\in \nat$ existing nodes randomly chosen with probability equal to their normalized degree at that moment. As for the SW type, its characteristic is that nodes are reachable to each other through short paths. To generate this structure, we use the Watts-Strogatz model \cite{Watts98}. This starts by arranging the $N$ nodes on a ring lattice, each connected to $m\in \nat$ neighbors, $m/2$ on each side. Then, with a fixed probability $p\in [0,1]$ for each edge, it decides to rewire it to a uniformly chosen node of the network.

We use a small population size of $N=100$ individuals, which however is sufficient for our demonstration. By rescaling the epidemic parameters, the same phenomena can be reproduced for larger networks. Note that the model parameters to obtain each used network, are mentioned explicitly in the associated figures.

\inlinetitle{Diffusion process and score-based DRA}{.}
We introduce a continuous-time SIS Markov process \cite{van2009virus,VanMieghem2011} into a given network, which we simulate at the node-level. For node $i$, the possible state transitions at time $t$ are: $X_{i,t}: 0 \rightarrow 1$ at rate $\beta \sum_j M_{ij} X_{j,t}$, and $X_{i,t}: 1 \rightarrow 0$ at rate $\rho R_{i,t}$, where $\beta$ is the contribution of an edge to the infection rate, and $\rho$ is the contribution of a received treatment to the node recovery rate. We set a fixed budget of $b=5$ resources. From the above SIS formulation we have dropped the self-recovery in order to emphasize the role of the compared DRA strategies.

Since in this paper our purpose is not to investigate the role of the different possible scoring functions, in the simulations we use a simple yet efficient function called \emph{Largest Reduction in Infectious Edges} (LRIE) \cite{Scaman15}. For each infected node, LRIE computes the difference in number between its neighbors that are healthy to the rest that are infected; formally: $S_{i,t} = \sum_j \left(M_{ij} \bar{X}_{j,t} - M_{ji} X_{j,t} \right)$. LRIE is greedy and dynamic, since node scores change when nodes' infection state and/or the network structure changes.

\subsection{Results}

\begin{figure}[t]
\centering
\vspace{-1mm}
\hspace{-3mm}
\subfigure[{\scriptsize Cutoff-based \SDRA on SW.}]
{
\clipbox{1.5pt 0pt 0pt 0pt}
{\includegraphics[width = 0.50\linewidth, viewport=10 10 553 530, clip]{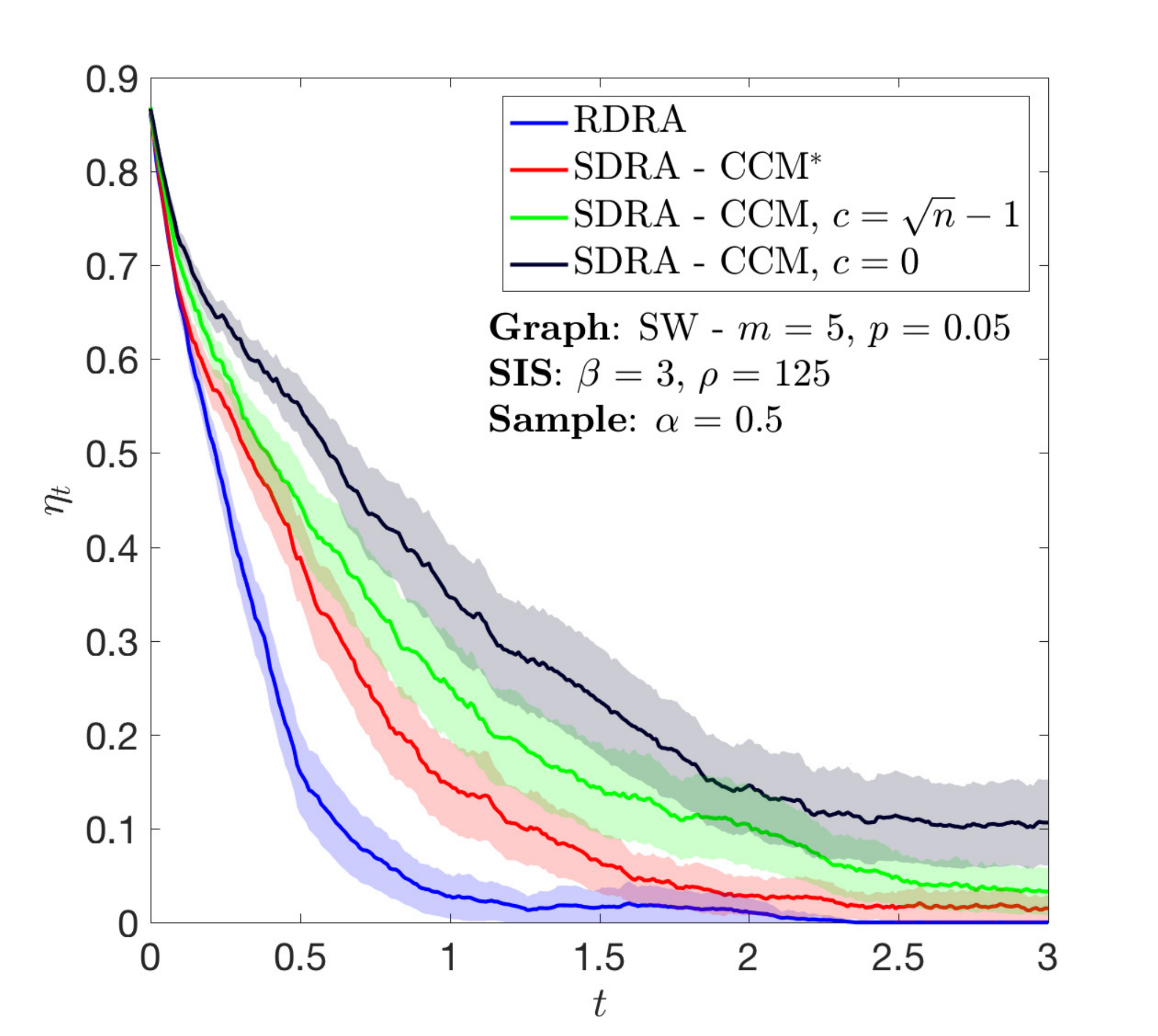}}\label{fig:sw_perc_inf_left}}
\vspace{2mm}
\hspace{-1mm}
\subfigure[{\scriptsize Threshold-based \SDRA on SW.}]
{
\clipbox{7pt 0pt 0pt 0pt}
{\includegraphics[width = 0.50\linewidth, viewport=10 10 553 530, clip]{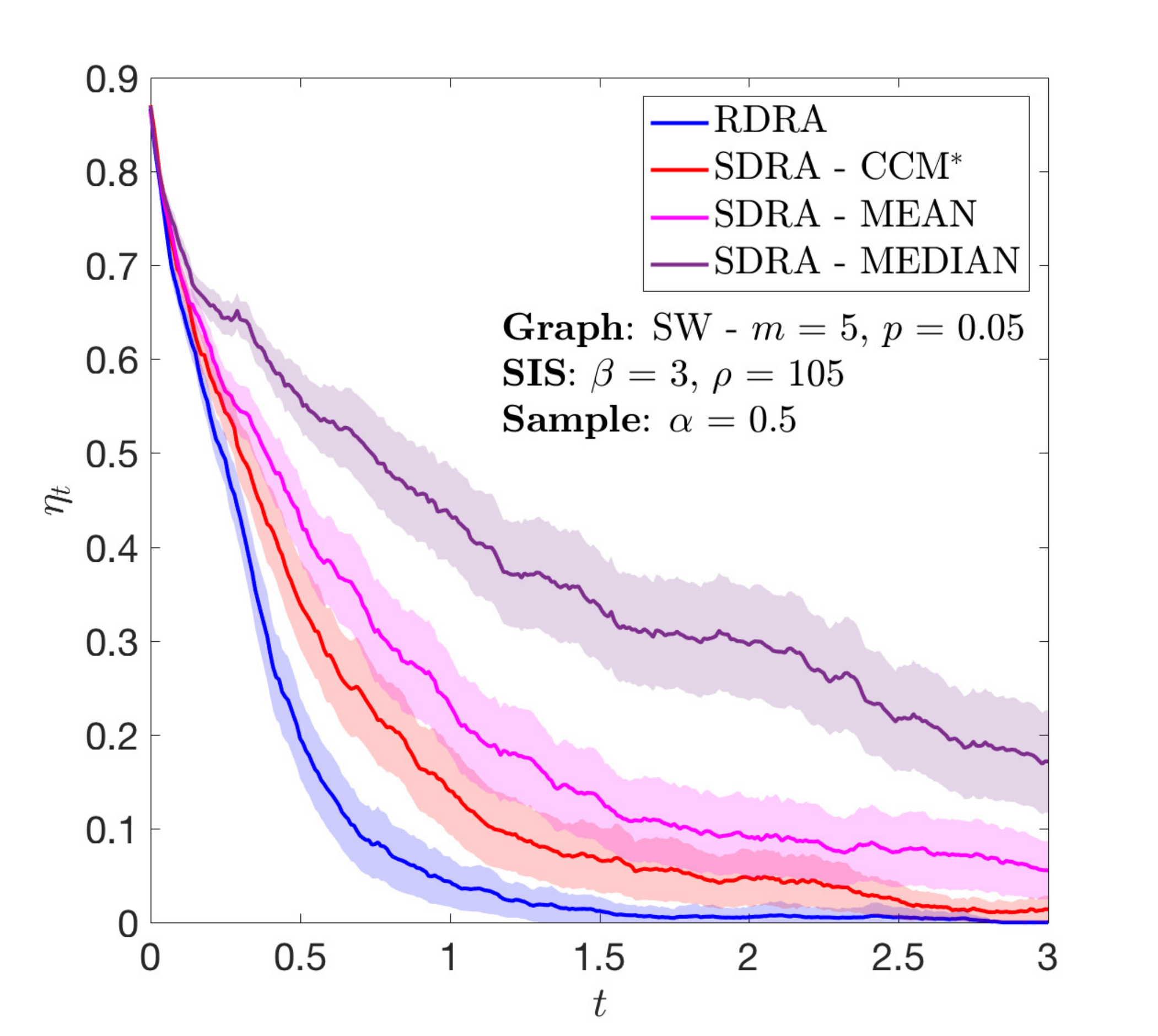}}\label{fig:sw_perc_inf_right}}\\
\vspace{-3mm}
\hspace{-3mm}
\subfigure[{\scriptsize Cutoff-based \SDRA on SF.}]{
\clipbox{1.5pt 0pt 0pt 0pt}
{\includegraphics[width = 0.5\linewidth, viewport=10 10 553 530, clip]{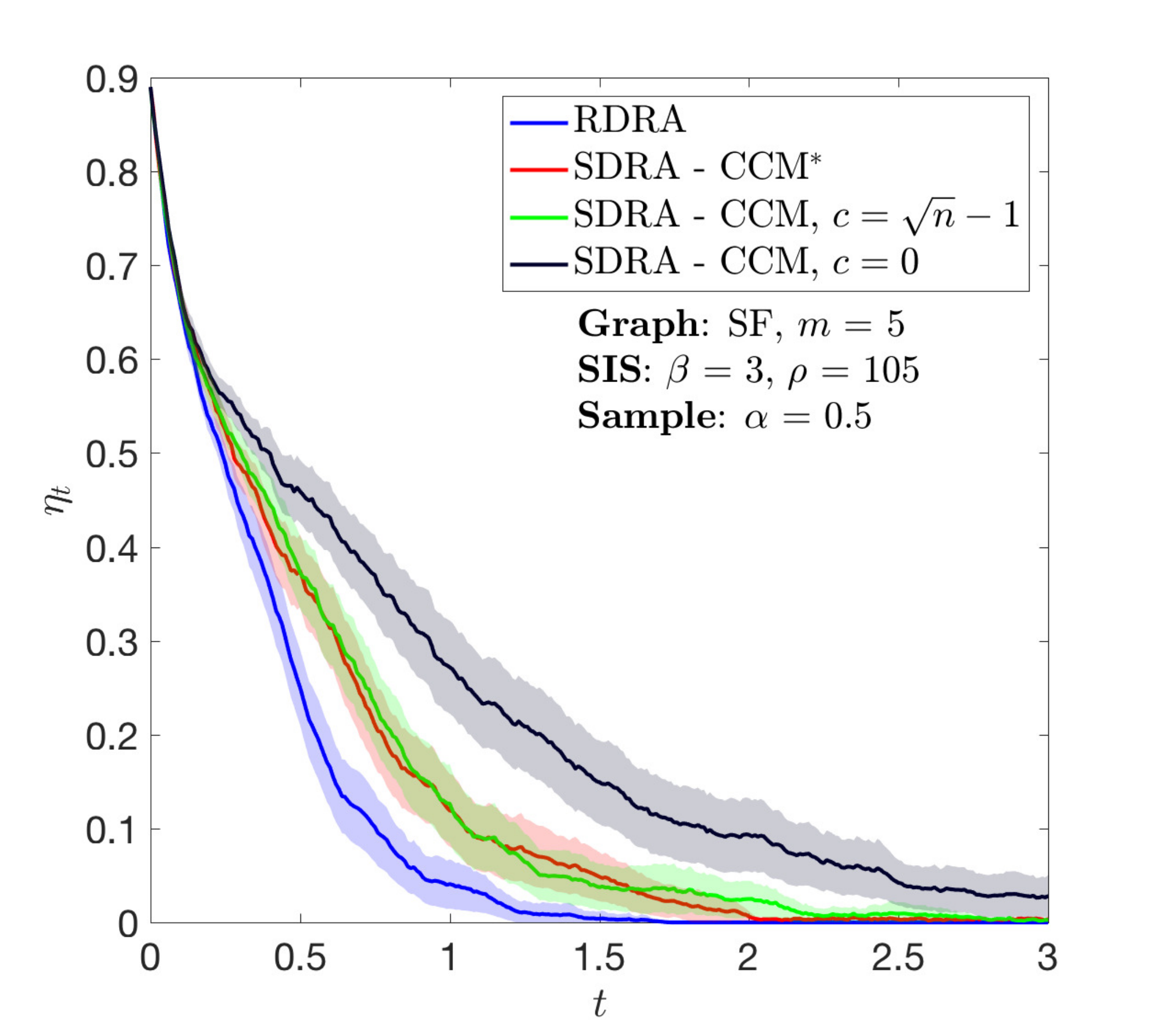}}\label{fig:pa_perc_inf_left}}
\subfigure[{\scriptsize Threshold-based \SDRA on SF.}]{
\clipbox{7pt 0pt 0pt 0pt}
{\includegraphics[width = 0.5\linewidth, viewport=10 10 553 530, clip]{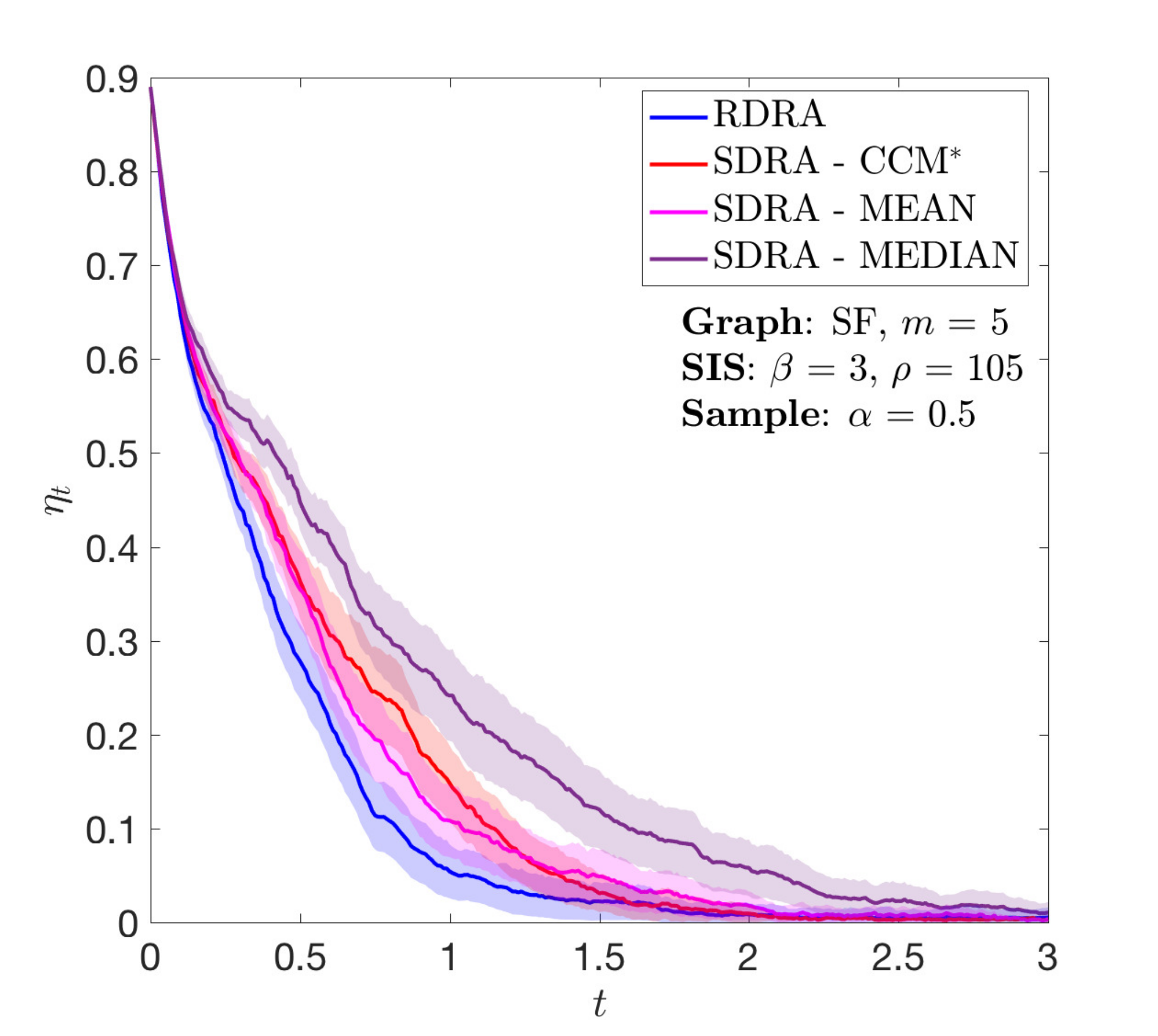}}\label{fig:pa_perc_inf_right}}\\
\caption{Comparison of cutoff-based and threshold-based SDRA strategies. The Restricted DRA is shown for reference; for the same reason, the proposed SDRA-CCM is also repeated in the right subfigure of each row. Average percentage of infected nodes $\eta_t$ through time for SW (top row) and for SF (bottom row) networks.}
\label{fig:perc_inf}
\end{figure}
\begin{figure}[t]
\centering
\hspace{-3mm}
\subfigure{
\clipbox{3pt 0pt 0pt 3.5pt}
{\includegraphics[width = 0.51\linewidth, viewport=2 5 560 540, clip]{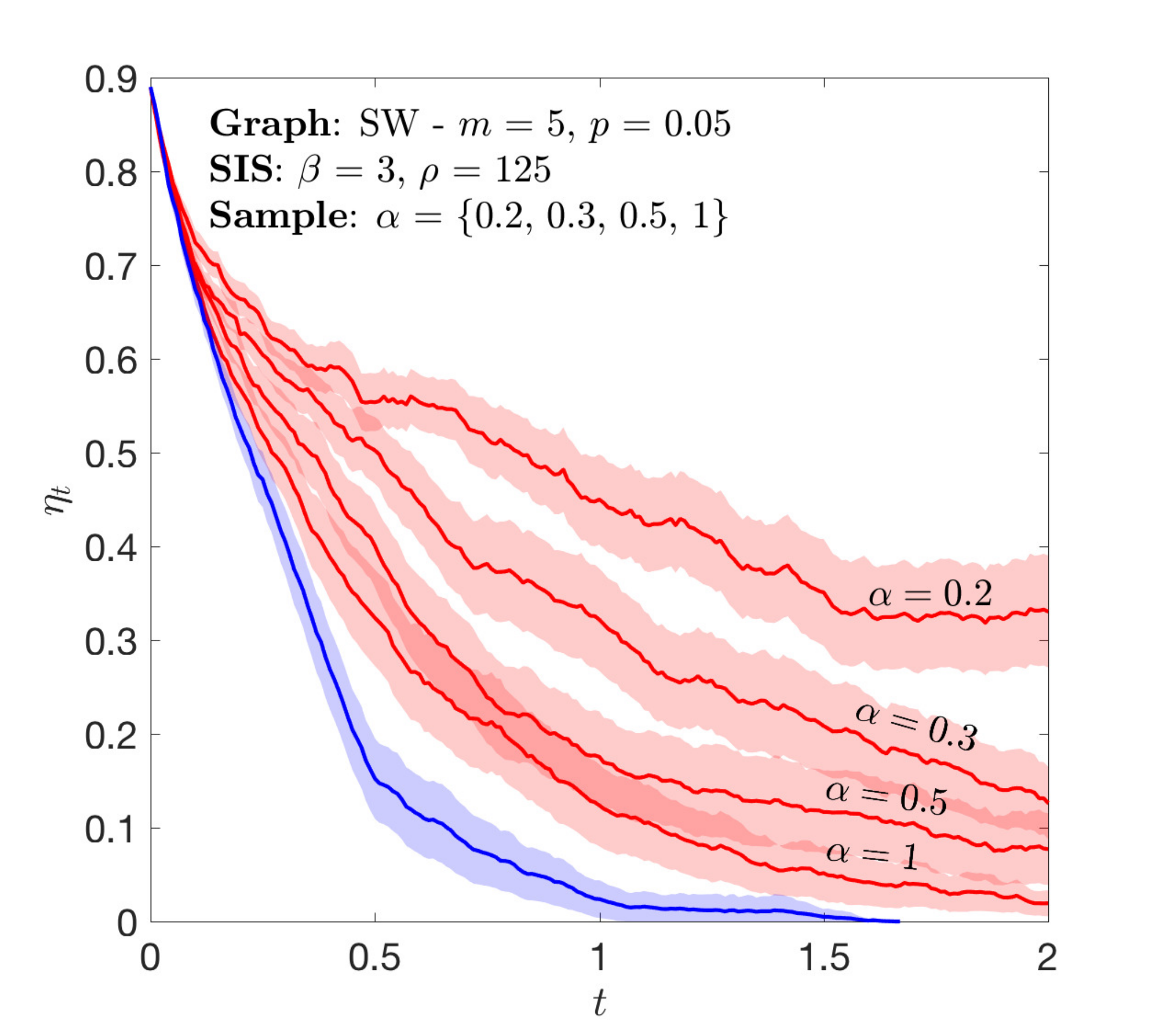}}\label{fig:sample_size_SW}}
\hspace{-1mm}
\subfigure{ 
\clipbox{8pt 0pt 0pt 3.5pt}
{\includegraphics[width = 0.51\linewidth, viewport=2 5 560 540, clip]{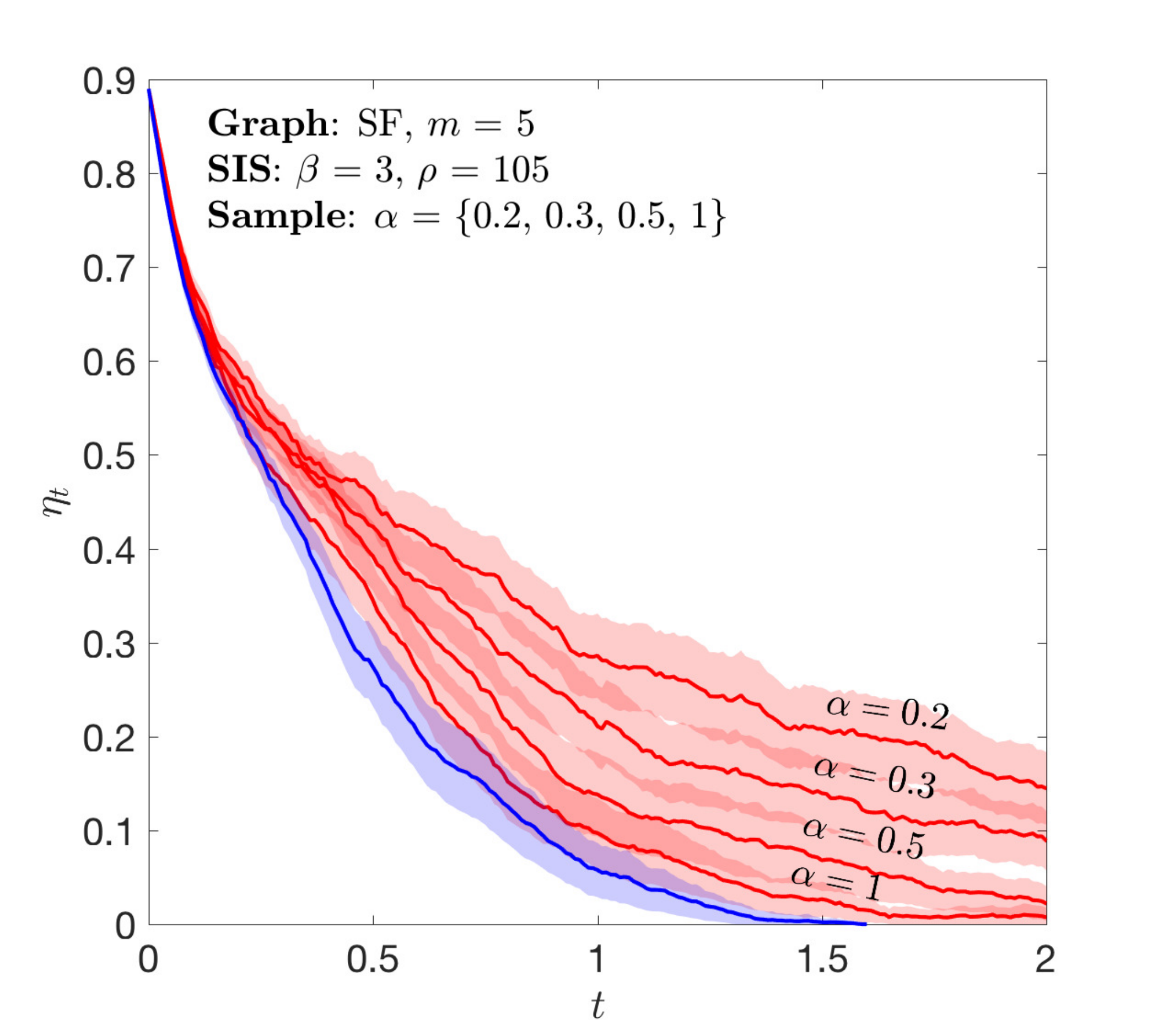}}\label{fig:sample_size_PA}}\\ 
\vspace{-2mm}
\caption{The average percentage of network infection though time using the sequential \CCMstar strategy (red lines), for various fixed sample sizes. The blue curves display the associated non-sequential \RDRA strategy with full access to nodes at each round (\ie $\alpha=1$).}
\label{fig:sample_size}
\end{figure}

\vspace{-1mm}
In the empirical study, our aim is to compare the performance of several DRA strategies $\policybold$, that follow \Alg{alg:sdra}. 
The offline selection strategy that picks the reachable candidates with the highest scores, is always plotted with blue curve as reference (see Fig.~\ref{fig:perc_inf}, \ref{fig:sample_size}). At each round, a fraction $\alpha\in[0,1]$ of the infected nodes becomes accessible to the \DM, $n_t = \lfloor \alpha \sum_iX_{i,t}\rfloor$, which is uniformly sampled from the population, \ie we set $\Lambda(n,\Xbold{})$ to be $\mathcal{U}(\Rankset{n}{\nodes})$. 

\inlinetitle{Cutoff-based \vs threshold-based strategies}{.}
\Fig{fig:perc_inf} displays the average percentage of infected nodes $\eta_t$ \wrt time $t$ using the compared DRA strategies on the two network types discussed earlier.
We start with the SW type at the top row, where on the left appears the cutoff-based \CCM strategy with various cutoffs, and on the right variations of the threshold-based strategy, \MEAN and \MEDIAN. In both subfigures, \CCMstar (red curve) is clearly the best performing approach. Also, here \MEAN is a lot better than \MEDIAN. However, in the respective simulations on a SF network (bottom row), the curves appear to be closer together; \MEAN and \MEDIAN have no more difference in performance. The \CCM is still better, but \CCMstar shows no improvement over the use of the simpler cutoff $c = \sqrt{n} - 1$.

\begin{figure}[t]
\centering
\hspace{-3mm}
\subfigure[{\scriptsize \RDRA on SW.}]{
\clipbox{1pt 0pt 0pt 3.5pt}
{\includegraphics[width = 0.5\linewidth, viewport=1 5 561 541, clip]{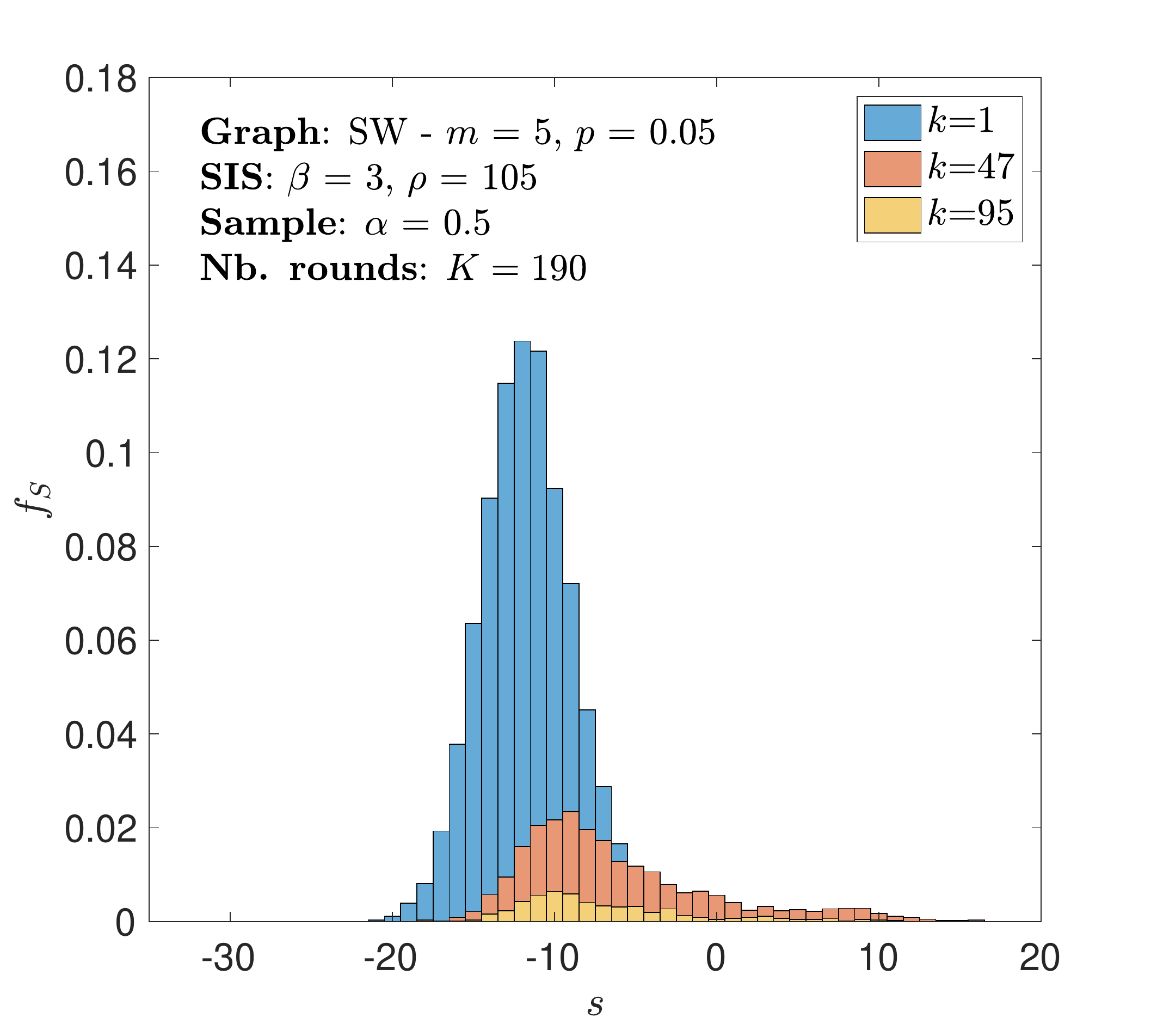}}\label{fig:sw_score_distrib_left}}
\subfigure[{\scriptsize \SDRA, \CCMstar on SW.}]{ 
\clipbox{7pt 0pt 0pt 3.5pt}
{\includegraphics[width = 0.5\linewidth, viewport=1 5 561 540, clip]{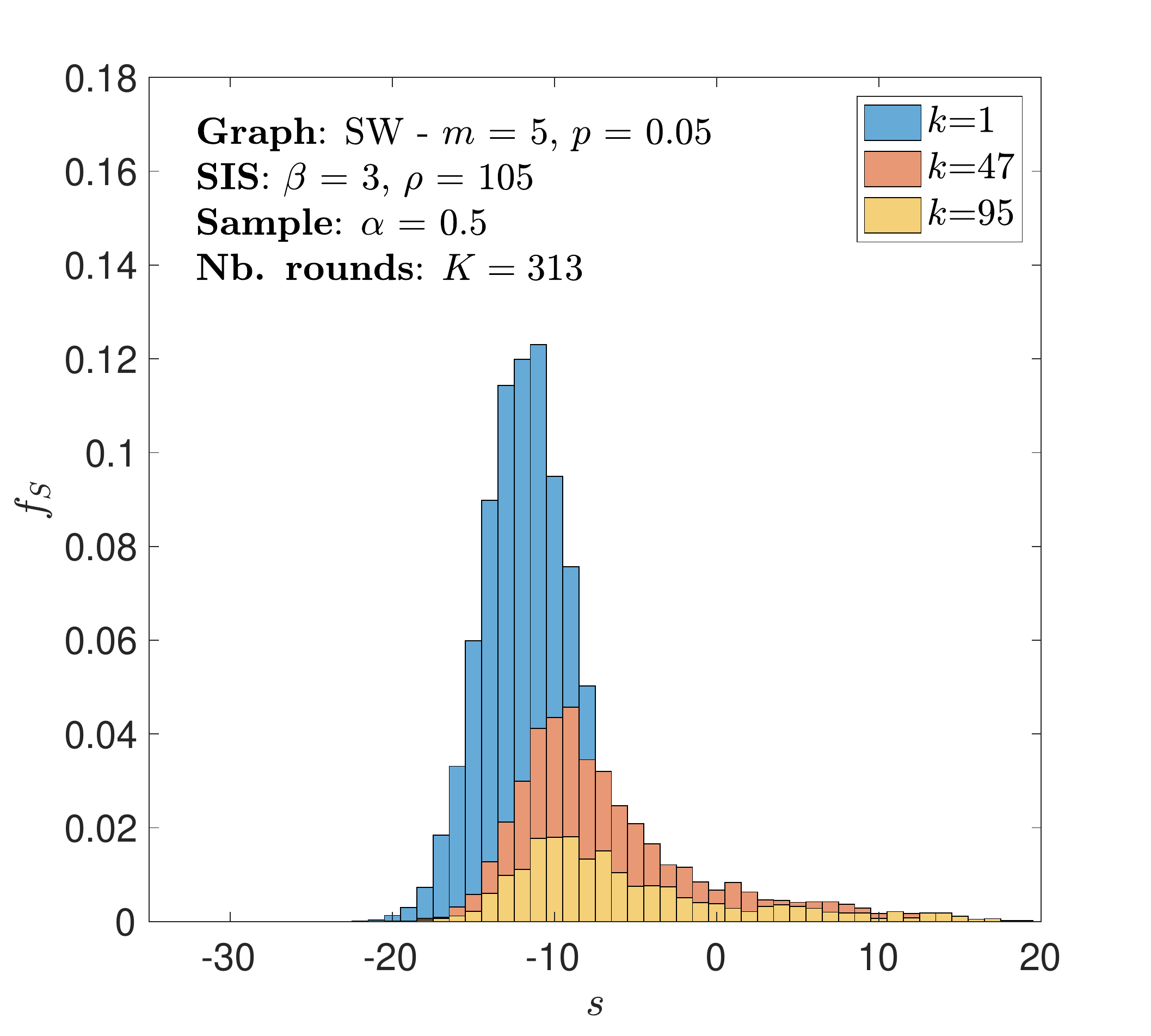}}\label{fig:sw_score_distrib_right}}\\
\vspace{-1mm}
\hspace{-2.6mm}
\subfigure[{\scriptsize \RDRA on SF.}]{
\clipbox{0.8pt 0pt 0pt 3.5pt}
{\includegraphics[width = 0.5\linewidth, viewport=1 5 561 540, clip]{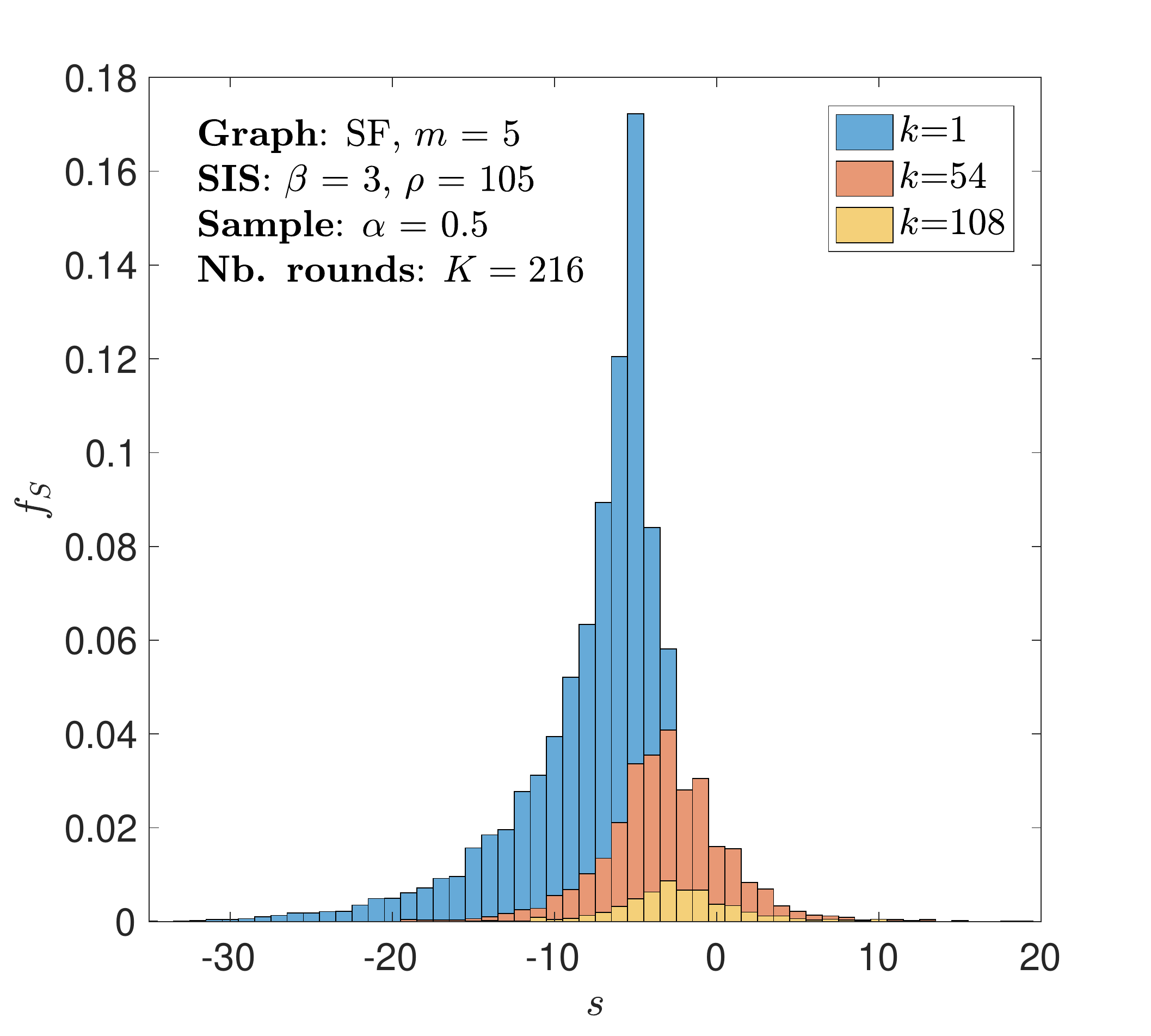}}\label{fig:sf_score_distrib_left}}
\subfigure[{\scriptsize \SDRA, \CCMstar on SF.}]{ 
\clipbox{7pt 0pt 0pt 3.5pt}
{\includegraphics[width = 0.5\linewidth, viewport=1 5 561 540, clip]{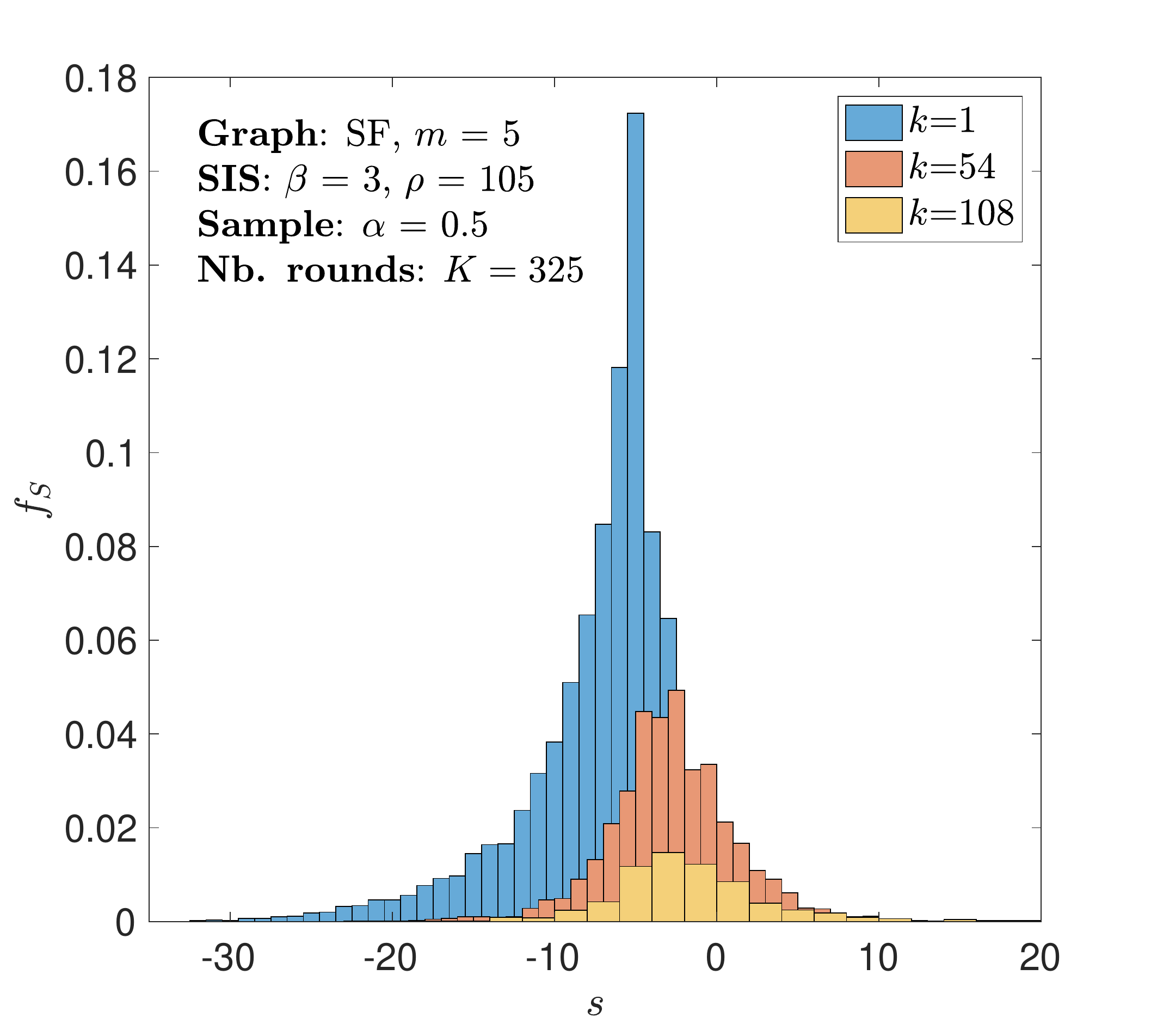}}\label{fig:sf_score_distrib_right}}
\vspace{-2mm}
\caption{Empirical \pdf $D(s)$ of the node scores, at different points in time, \ie different rounds $\t$, for SW (top row) and for SF (bottom row) networks, when applying \RDRA or \SDRA (the \CCMstar). 
}
\label{fig:score_distrib}
\end{figure}

To investigate the behavior of strategies further, in \Fig{fig:score_distrib} we plot the score distribution $D(s)$ (here, this comes from LRIE) for all the infected nodes of the network, at three different rounds (\ie time instances) shown in different colors. The top and bottom rows refer respectively to SW and SF networks. \Fig{fig:sw_score_distrib_left} and \Fig{fig:sf_score_distrib_left} show the $D(s)$ obtained using a \RDRA strategy (blue curve in \Fig{fig:sw_perc_inf_left}), while for \Fig{fig:sw_perc_inf_right} and \Fig{fig:pa_perc_inf_right} the sequential strategy used is the \CCMstar (red curve in \Fig{fig:sw_perc_inf_left}). Starting from almost identical $D(s)$ per row at $k=1$ (initialization with the same infection level), we observe that throughout the rounds the difference between the distributions of the \RDRA and \SDRA strategies is larger for SW networks. This is as expected, since in that example the two strategies have larger difference in performance. We also observe that, in the case of SF networks, the $D(s)$ leans towards a Gaussian-like shape, which explains why \MEAN and \MEDIAN behave similarly, contrary to the more skewed shape obtained for a SW network.

Something easy to see in these simulations is that the network structure plays a crucial role in how the epidemic spreads and sets the difficulty level to a strategy that tries to contain it. Also, the highly evolving shape of the score distribution throughout the course of rounds illustrates the challenges that SDRA strategies need to address in order to be sufficiently effective. 

\inlinetitle{Sample size}{.} As described, the sampling is performed on the infected nodes and so far we used a fixed ratio, $\alpha=0.5$. To analyze the impact of the sample size on the efficiency of an CCM strategy, we plot in \Fig{fig:sample_size} the average percentage of infected nodes \wrt time for various sampling ratios. We observe that the SDRA is less sensitive to the sample size on SF networks (right) than on SW networks (left). Furthermore, regardless to the network structure, increasing the sample size does not improve linearly the efficiency of the algorithm.

\section{Conclusion and discussion}\label{sec:conclusion}
This study aimed towards bringing \DRAfull (\DRA) strategies closer to real-life constraints. We reviewed their strong assumption that the \DM has full information and access to all network nodes, at any moment a round of decisions takes place: anytime needed, she can instantaneously reallocate resources to any nodes indicated by a criticality scoring function. 
We significantly relaxed this assumption by first introducing the \emph{\RDRAshort} model, where only a sample of nodes becomes accessible at each round of decisions. Inspired by the way decisions are taken while care-seekers appear at a healthcare unit, we next proposed the \emph{Sequential \DRA} model that limits further the control strategy so as to have sequential access to only a sample of nodes selected at each round. This setting offers a completely new perspective to the dynamic \DP control: the \DM examines the nodes one-by-one and decides immediately and irrevocably whether to treat them or not by reallocating treatment resources. This online problem is put in relation with recent work in the Sequential Selection literature where efficient algorithms have been presented for the selection of items from a sequence for which little or no information is available in advance. Special mention should be made to the \emph{\MSSPfull} (\MSSP) that has been found to be particularly fitting for handling the sequential reallocation of resources. Finally, according to our simulations on SIS epidemics, where we compared the performance of several variants of the above \DP control models, we conclude that the cutoff-based \CCMstar is a very promising approach for the setting of sequential \DP control.

\bibliographystyle{abbrv} 
\bibliography{SDRA_2019_arxiv}

\end{document}